# Spin wave modes in a cylindrical nanowire in crossover dipolar-exchange regime

J. Rychły[1,2,*], V. S. Tkachenko[3], J. W. Kłos[1,4], A. Kuchko[5,6], M. Krawczyk[1]

[1] *Faculty of Physics, Adam Mickiewicz University in Poznań, Umultowska 85, Poznań, Poland*
[2] *Materials Science Division, Argonne National Laboratory, Argonne, IL, United States*
[3] *Vasyl` Stus Donetsk National University, 600-richya str. 21, Vinnytsia, 21021, Ukraine*
[4] *Institute of Physics, Greifswald University, Felix-Hausdorff-Str. 6, 17489 Greifswald, Germany*
[5] *Institute of Magnetism of NAS of Ukraine, Ukraine*
[6] *Igor Sikorsky Kyiv Polytechnic Institute, Kyiv, Ukraine*

Nanoscale magnetic systems have been studied extensively in various geometries, such as wires of different cross-sections, arrays of wires, dots, rings, etc. Such systems have interesting physical properties and promising applications in advanced magnetic devices. Uniform magnetic nanowires are the basic structures which were broadly investigated. However, some of their dynamical properties, like: (anti)crossing between the spin wave modes and impact of the magnetic field on spin wave spectrum, still need to be exploited. We continue this research by investigation of the spin wave dynamics in solid Ni nanowire of the circular cross-section. We use two approaches: semi-analytical calculations and numerical computations based on finite element method. We solve coupled Landau-Lifshitz and Maxwell equations and consider both magnetostatic and exchange interactions. We identify the dispersion brunches and its (anti)crossing by plotting the spatial profiles of spin wave amplitudes and magnetostatic potential. We also check how we can tune the spectrum of the modes by application of the external magnetic field and how it affects the modes and their dominating type of interaction.

## 1. Introduction

The studies on magnetization dynamics in ferromagnetic systems began in the middle of the twentieth century [1,2] and are still active due to interest in the application of spin waves and spin currents to data processing [3,4] and data storage [5,6] performed with the use of magnetic nanostructures [7,8,9,10]. The fundamental characteristic of wave excitation is dispersion relation. The dispersion of the spin waves can be shaped by the geometry of the system. In the confined geometries (films, wires, and dots) new effects appear which affect both magnetization configuration and also the dynamical coupling of the magnetic moments. The presence of surfaces influences directly (by modes quantization, induction of the dynamical demagnetizing field) or indirectly (by the formation of "magnetic landscape" determining the static magnetic configuration and the static internal field) on spin wave dynamics.

The possible stable or metastable magnetic configurations are determined by bulk interactions (exchange stiffness, static magnetization, bulk magnetocrystalline anisotropy), state of the surfaces (surface magnetocrystalline anisotropy) and by the geometry of the system [11,12]. The latter factor can be designed in the fabrication process. The magnetic configuration can be further controlled by the application of the external bias (magnetic field, temperature gradient, electric field or strain in multiferroic systems) [13,14,15]. This allows using magnetic nanostructures (arrays of wires or dots) of a specific structure for high-density information storage, where the spatial size of information bit is determined by the size of the nanoelement [16,17]. However, in more advanced design even a single magnetic nanoelement (nanowire) can store the sequence of bits. This idea was implemented in the racetrack memories where the bits stored in nanowire are coded in domain walls [18]. This explains extensive research on static magnetic properties in confined geometries in last years.

---
[*] Electronic address: rychly@amu.edu.pl

The adjustment of parameters in films or wire system [19,20] can be used also for shaping spin wave propagation [21,22,23,24,25]. The effect of confinement for exchange waves (spin waves determined by the exchange interactions) results mostly in the quantization of spin wave modes. In dipolar regime, the static and dynamic demagnetizing fields induced by the presence of surfaces/interfaces can introduce additional effects (surface localization, anisotropy, and nonreciprocity of propagation), which can be further tuned by the external bias. The increase of the magnitude of the external magnetic field leads not only to the shift of the dispersion branches to higher frequencies (as it is for the exchange waves) but also to the change of the group velocity of spin waves. The nanostructures in form of the regular planar arrays of weakly interacting magnetic nanoelements (e.g. dots or wires) combine the features of confined and propagating modes – the spin wave dispersion in such systems follows the trend of the dispersion of a homogeneous film, but split into relatively narrow bands resulting from partial confinement in nanoelements [26,27].

Our study is motivated by potential application of magnetic nanowires for spin wave transmission and shaping of spin waves signals in magnonic system. In future integrated magnonic devices and systems, the wires seem to be important not only as passive elements for spin wave transmission but can be also used for spin wave processing (filtering [28], tuning group velocity [29], non-reciprocal transmission [30]). Therefore the in-depth studies of spin wave propagation in magnonic nanowires are essential for designing magnonic systems. Although a uniform magnonic wire is a basic structure, the spin wave spectrum of such a system is complicated even for a saturated magnetic configuration.

The thin nanowires of rectangular cross-section (produced be top-down lithographic techniques) are relatively simple in theoretical analysis, as far as the system can be treated as a quasi-2D structure with approximately homogeneous spin wave profile across the thickness of the nanowire. In this approach, the quantization of the modes of the waveguide results only from the finite (in-plane) width of the wire [31]. The circular cross-section is natural shape for the nanowires fabricated based on bottom-up approach, like in pores of the anodic aluminium oxide templates [32] or by fully chemical methods, e.g. by reduction of chloride hexahydrate of ferromagnetic metal in a magnetic field. The calculation of spin wave dispersion relation in the ferromagnetic wire was done in early 60-ties [33]. The authors considered purely dipolar waves in cylindrical ferromagnetic nanowire magnetized along its axis. They found that all modes have negative group velocities in the whole range of the wave number and can be grouped into two ranges of frequencies: lower range – in which modes are quantized (are oscillating) both in radial and azimuthal directions and higher range – in which modes are quantized only in azimuthal direction and localized in radial direction at the surface of the wire. The modes localized on the surface exist only for the wave numbers smaller than some critical value, which depends on the azimuthal number of the mode. The spin wave spectrum of this system shows some similarities to the spectrum of the in-plane magnetized layer [34,35,36].

The theoretical investigations of spin wave dynamics in the cylindrical nanowire, with both dipolar and exchange interactions taken into account, were reported in Ref. [37]. This investigation, based on the continuum model, was then extended to the case of the wire of arbitrary cross-section [38]. Transversely magnetized flat wires were studied in [39] for the range of wire widths and thicknesses, a quantitative description of the spin wave eigenmode frequencies and spatial profiles as a function of the wire width was provided.

The increase of the interest of spin wave propagation in cylindrical nanowires operating in crossover dipolar-exchange regime results from the development of fabrication methods, experimental techniques for characterization of spin wave dynamics and numerical tools. The works of Arias and Mills [37,38] were supplemented by experimental studies (BLS measurements) [40] and the calculations based on spin models [41]. The continuous model introduced by Arias and Mills [37] was used later to discuss the spin wave dispersion in magnetic nanorods and nanotubes [42,43,44]. We can find also further experimental and theoretical works concerning the spin wave dynamics in nanowires of non-circular cross-section or under the influence of magnetic field applied in a direction which is not parallel to the (easy) axis of the wire [45,46,47,48,49] or magnetization-modulated cylindrical nanowires [50].

Although all these extensive studies, there are still issues needed to be explained for cylindrical nanowires operating in the crossover dipolar-exchange regime. In our paper, through semi-analytical and numerical calculations of the spin wave spectra in an infinitely long nanowire in the magnetically saturated state, we are going to address the following:
(i) trace the evolution of the spin wave modes originating from the fundamental mode with increasing wave number by examination of its profiles; particular attention will be paid on the interaction with other modes,
(ii) study contribution of dipolar and exchange interactions to the spin wave spectrum and its evolution with increased external magnetic field,
(iii) explain the mode dependent group velocity and the influence of the external bias magnetic field on shaping spin wave dispersion relation.

We use two approaches in our studies: analytical calculations and numerical computations based on the finite element method (FEM). In both techniques, we solve continuous model described by linearized Landau-Lifshitz equation (LLE) and Maxwell equations (ME) with dipolar and exchange interactions taken into account. On the surfaces of the wire, we use natural boundary conditions resulting from the ME.

The manuscript is organized as follows. In the next section, we describe in details the structures which we are going to investigate, then we present the derivations concerning the analytical model and outline of computational technique we use. In the section *'Results'* we show and discuss the outcomes of analytical and numerical studies, which are summarized in the section '*Conclusions*'.

## 2. The analytical and numerical models

We present a systematic approach to analyzing spin wave spectrum in uniform magnetic nanowire of the finite cross section. We use a continuum medium theory of the dispersion of dipolar-exchange spin waves in magnetic nanowire of the circular cross-section. It can be described as an infinite cylinder with radius $R$. The martial parameters: gyromagnetic ratio $\gamma$, saturation magnetization $M_0$ and exchange constant $A$ are assumed to be constant throughout the cylinder. The easy magnetization axis and the direction of the applied magnetic field are both parallel to the axis of the nanowire.

To describe magnetization dynamics we solved the LLE which is the equation of motion for the magnetization vector $\mathbf{M}(\mathbf{r},t)$:

$$\frac{d\mathbf{M}(\mathbf{r},t)}{dt} = \gamma\mu_0[\mathbf{M}(\mathbf{r},t) \times \mathbf{H}_{eff}(\mathbf{r},t)], \tag{1}$$

where: $\mu_0$ is the permeability of vacuum, $\gamma$ denotes gyromagnetic ratio and $\mathbf{H}_{eff}$ is the effective magnetic field. The term on the right hand side is a torque which describes the precessional motion of the magnetization around the direction of the effective magnetic field, the damping is neglected.

The effective magnetic field $\mathbf{H}_{eff}$ can consist of many terms, but in this paper, we will consider the external magnetic field $\mathbf{H_0}$, exchange field $\mathbf{H}_{ex}$ and dipolar field $\mathbf{H}_m$: $\mathbf{H}_{eff}(\mathbf{r},t) = \mathbf{H_0} + \mathbf{H}_{ex}(\mathbf{r},t) + \mathbf{H}_m(\mathbf{r},t)$.

We assumed that the static magnetic configuration of the considered nanowire is saturated along the axis of the nanowire (i.e. along the easy axis of the system). First, we studied the spin wave dynamics in the absence of an external magnetic field and then we investigated how the presence of the external magnetic field affect the results.

The investigated nanowire has a cylindrical symmetry. Therefore, it is reasonable to use the cylindrical coordinates system $(\rho, \phi, z)$ where: the coordinate $z$ is a distance along the axis of the wire, the symbol $\rho$ denotes radial distance from the $z$-axis and the angle $\phi$ is the azimuthal angle, marked in the plane perpendicular to the axis of the wire.

In the linear regime, the magnetization vector in the nanowire can be described as $\mathbf{M}(\mathbf{r},t) = \hat{\mathbf{z}}M_0 + \mathbf{m}(\mathbf{r},t)$, where the $\hat{\mathbf{z}}$ is a unit vector along the wire. We assume here the uniform and saturated ground magnetic state: $\hat{\mathbf{z}}M_0$. Dynamic part of the magnetization vector is small in reference to the magnetization $M_0$: $|\mathbf{m}| \ll M_0$, thus $M_0$ is assumed to equal to the saturation magnetization. Both components $m_\rho(\mathbf{r},t)$ and $m_\phi(\mathbf{r},t)$ of the dynamic part of magnetization $\mathbf{m}(\mathbf{r},t)$ oscillate harmonically in time $\mathbf{m}(\mathbf{r},t) = \left(\hat{\boldsymbol{\rho}}\, m_\rho(\mathbf{r}) + \hat{\boldsymbol{\theta}}\, m_\theta(\mathbf{r})\right)e^{-i\omega t} = \mathbf{m}(\mathbf{r})e^{-i\omega t}$, when the magnetization $\mathbf{M}(\mathbf{r},t)$ precesses around the $z$-axis.

For considered magnetic configuration and the geometry of the structure, the static components of the exchange field and the demagnetizing field are equal to zero. Therefore, the exchange field is defined as: $\mathbf{H}_{ex}(\mathbf{r},t) = \alpha \nabla^2 \mathbf{m}(\mathbf{r})e^{-i\omega t}$, where the parameter $\alpha$ is a squared exchange length and is related to the magnetization saturation $M_0$ and exchange constant $A$ in the following way: $\alpha = \frac{2A}{\mu_0 M_0^2}$. The amplitude $\mathbf{h}_m(\mathbf{r})$ of the demagnetizing field: $\mathbf{H}_m(\mathbf{r},t) = \mathbf{h}_m(\mathbf{r})e^{-i\omega t}$ can be related to the magnetostatic potential $\varphi(\mathbf{r})$: $\mathbf{h}_m(\mathbf{r}) = -\nabla\varphi(\mathbf{r})$ induced by the precessing magnetization: $\nabla^2 \varphi(\mathbf{r}) = \nabla \cdot \mathbf{m}(\mathbf{r})$. The above relation between demagnetizing field and magnetization can be derived from ME using magnetostatic approximation [51]. The effective magnetic field inside the nanowire will be described as:

$$\mathbf{H}_{eff}(\mathbf{r},t) = \mathbf{H_0} + \alpha \nabla^2 \mathbf{m}(\mathbf{r})e^{-i\omega t} - \nabla\varphi(\mathbf{r})e^{-i\omega t}. \tag{2}$$

For calculations of infinitely-long 60 nm radius Ni nanowire, we take material parameters from [42]: saturation magnetization $M_{0,\,\text{Ni}} = 0.48 \times 10^6 \frac{\text{A}}{\text{m}}$, exchange constant: $A_{\text{Ni}} = 7.46 \times 10^{-13} \frac{\text{J}}{\text{m}}$ (we chose the underestimated value $A_{\text{Ni}}$ to ensure the comparison of our outcome to the results presented in Ref. [40]), and the gyromagnetic ratio $\gamma = 193.6 \frac{\text{GHz}}{\text{T}}$. We considered the strength of the external magnetic applied field $\mu_0 H_0$ (if applied) equal to 100 mT.

## 2.1. Analytical model

By using the method introduced in Refs. [37,38], we can obtain the characteristics describing the dynamic of the magnetic system: magnetostatic potential, dipolar field and dynamic magnetization. Because the investigated nanowire has cylindrical symmetry, we are looking for the solutions in the form of cylindrical harmonics with a plane wave solution in the $z$-direction (propagating with wave number $k$) and quantized in the azimuthal direction (with the quantum number $q = 0, \pm 1, ...$): $f(\rho, \phi, z, t) \propto F_q(\kappa\rho)e^{i(q\phi+kz)}e^{i\omega t}$ [37]. The radial dependence, described by the factor $F_q(\kappa\rho)$, can be expressed in terms of the Bessel functions. The dimensionless argument of Bessel functions has the form $\kappa\rho$ where the parameter $\kappa$ can be considered as a wave number in the radial direction. The spatial profile of the radial factor $F_q(\kappa\rho)$ depends on the order of the Bessel functions (determined by the quantum number $q$) and the value of parameter $\kappa$. The last factor $e^{i\omega t}$ describes the temporal changes.

The wave number $k$ (for propagation along the $z$ – axis) and the parameter $\kappa$ (influencing the spatial oscillation in the radial direction) are depended on each other. This dependence can be found strictly for the eigenmodes of given frequency [37]. For the unconstrained system the magnetostatic potential accompanying the spin wave eigenmodes can be written in the following form in the cylindrical coordinate system: $\varphi(\mathbf{r}, t) = J_q(\kappa\rho)e^{i(q\phi+kz)}e^{i\omega t}$, where the symbol $J_q$ stands for the Bessel function of the first kind [52]. Substituting this solution to LLE (1) and using the relation between magnetization $\mathbf{m}(\mathbf{r}, t)$ and magnetostatic potential $\varphi(\mathbf{r}, t)$, derived from ME, we can obtain the following equation relating the values of $\kappa$ and $k$ [37]:

$$\alpha^2(\kappa^2 + k^2)^3 + \alpha(2H + 1)(\kappa^2 + k^2)^2 + (H(H + 1) - \alpha k^2 - \Omega)(\kappa^2 + k^2) - Hk^2 = 0. \tag{3}$$

We introduced in (3) two dimensionless parameters: $\Omega = \frac{\omega}{\omega_M}$ (where $\omega_M = \gamma\mu_0 M_0$) and $H = \frac{H_0}{M_0}$.

When we select the wave number $k$ (the free choice of $k$ is also valid for wire geometry) then the parameter $\kappa$ has to take defined values. The Eq. 3 is the third order equation in respect to $\kappa^2$ and for each freely selected value of $k$, we can obtain up to three corresponding values of the parameter $\kappa$. Therefore this parameter can be indexed by the integer $n=1, 2, 3$ and written as $\kappa_n$.

For a magnetic wire of the radius $R$, we have to distinguish two regions. Inside the wire, in the magnetic material ($\rho < R$), we accept the oscillatory solutions where the radial factor is the superposition of Bessel functions $J_q(\kappa_n\rho)$ for $n = 1, 2, 3$. In the non-magnetic surrounding of the nanowire ($\rho > R$), we look for a monotonously decaying solution as $\rho \to \infty$. Therefore, we chose for $\rho > R$ the modified Bessel function of the second kind $K_q(k\rho)$ as a radial factor:

$$\varphi_1(\mathbf{r}) = \sum_{n=1}^{3} A_n J_q(\kappa_n\rho) e^{iq\phi+ikz}, \quad \rho < R, \tag{4}$$

$$\varphi_2(\mathbf{r}) = B \cdot K_q(k\rho) e^{iq\phi+ikz}, \quad \rho > R.$$

Here the index $q$ is an order of the Bessel functions. The coefficients: $A_n$ and $B$ are unknown amplitudes which can be determined after applying the boundary conditions at $\rho = R$. The corresponding $\rho$ and $\phi$ components of the dynamic dipolar field $\mathbf{h}_{m,(1,2)} = \hat{\boldsymbol{\rho}}\, h_{m,\rho,(1,2)}(\mathbf{r}) + \hat{\boldsymbol{\varphi}}\, h_{m,\varphi,(1,2)}(\mathbf{r})$ read:

$$h_{m,\rho,1}(\mathbf{r}) = \sum_{n=1}^{3} \left[ A_n \left( \frac{q}{\rho} J_q(\kappa_n\rho) - \kappa_n J_{q+1}(\kappa_n\rho) \right) \right] \exp(i(q\phi+kz)), \quad \rho < R,$$

$$h_{m,\phi,1}(\mathbf{r}) = -\frac{i}{\rho} q \sum_{n=1}^{3} A_n J_q(\kappa_n\rho) \exp(i(q\phi+kz)), \quad \rho < R,$$

$$h_{m,\rho,2}(\mathbf{r}) = \frac{B}{\rho} \left( -k\rho \cdot K_{q+1}(k\rho) + qK_q(k\rho) \right) \exp(i(q\phi+kz)), \quad \rho > R, \tag{5}$$

$$h_{m,\phi,2}(\mathbf{r}) = -\frac{i}{\rho} qBK_q(k\rho) \exp(i(q\phi+kz)), \quad \rho > R.$$

The components of dynamic magnetization inside the nanowire, expressed in the cylindrical coordinate system take the form:

$$m_\rho(\mathbf{r}) = \frac{1}{2} \sum_{n=1}^{3} \left[ A_n \kappa_n \left( \frac{J_{q+1}(\kappa_n\rho)}{H + \alpha(\kappa_n^2 + k^2) + \Omega} + \frac{J_{q-1}(\kappa_n\rho)}{H + \alpha(\kappa_n^2 + k^2) - \Omega} \right) \right] \exp(i(q\phi+kz)),$$

$$m_\phi(\mathbf{r}) = \frac{i}{2} \sum_{n=1}^{3} \left[ A_n \kappa_n \left( \frac{J_{q+1}(\kappa_n\rho)}{H + \alpha(\kappa_n^2 + k^2) + \Omega} - \frac{J_{q-1}(\kappa_n\rho)}{H + \alpha(\kappa_n^2 + k^2) - \Omega} \right) \right] \exp(i(q\phi+kz)). \tag{6}$$

One needs to apply boundary conditions in order to determine four unknown amplitudes and to obtain dispersion relation. Our solutions are subjects to the boundary conditions for magnetic induction and field which require that the normal (or radial in the case of the cylindrical coordinate system) components of the magnetic induction $\mathbf{b} = \mu_0(\mathbf{h}_m + \mathbf{m})$ and tangential (or azimuthal in the cylindrical coordinate system) components of $\mathbf{h}_m$ are continuous across the surface of the coaxial wire. Firstly, from the continuity of tangential component of the magnetic field $h_\phi$ and normal component of induction $b_\rho$ on the wire surface $\rho = R$ we shall have:

$$h_{m,\phi,1}(\mathbf{r})\big|_{\rho=R} = h_{m,\phi,2}(\mathbf{r})\big|_{\rho=R} \quad \text{and} \quad b_{\rho,1}(\mathbf{r})\big|_{\rho=R} = b_{\rho,2}(\mathbf{r})\big|_{\rho=R}. \tag{7}$$

For dynamic magnetization components we use (in analytical calculations) the exchange boundary conditions in the form:

$$\left.\frac{\partial}{\partial \rho} m_{(\rho,\phi)}(\mathbf{r})\right|_{\rho=R} = 0, \tag{8}$$

according to Ref.[1]. The system of equations which is formed from (7) and (8) relations is linear with respect to the unknown amplitudes and has non-trivial solutions if and only if its determinant $(\det(\hat{D}))$ is equal to zero:

$$\begin{pmatrix} J_q(\kappa_1 R) & J_q(\kappa_2 R) & J_q(\kappa_3 R) & -K_q(k\,R) \\ F_q(\kappa_1 R) & F_q(\kappa_2 R) & F_q(\kappa_3 R) & \frac{kR \cdot K_{q+1}(kR) - qK_q(kR)}{R} \\ P'_{q,+}(\kappa_1 R) & P'_{q,+}(\kappa_2 R) & P'_{q,+}(\kappa_3 R) & 0 \\ P'_{q,-}(\kappa_1 R) & P'_{q,-}(\kappa_2 R) & P'_{q,-}(\kappa_3 R) & 0 \end{pmatrix} \cdot \begin{pmatrix} A_1 \\ A_2 \\ A_3 \\ B \end{pmatrix} = \hat{D} \cdot \begin{pmatrix} A_1 \\ A_2 \\ A_3 \\ B \end{pmatrix} = 0. \tag{9}$$

Here we introduced the notations:

$$F_q(\kappa_n R) = \frac{q}{R} J_q(\kappa_n R) - \kappa_n J_{q+1}(\kappa_n R) + 2\pi\kappa_n P_{q,+}(\kappa_n \rho),$$

$$P_{q,\pm}(\kappa_n \rho) = \frac{J_{q+1}(\kappa_n \rho)}{H + \alpha(\kappa_n^2 + k^2) + \Omega} \pm \frac{J_{q-1}(\kappa_n \rho)}{H + \alpha(\kappa_n^2 + k^2) - \Omega}, \tag{10}$$

$$P'_{q,\pm}(\kappa_n R) = \left.\frac{\partial P_{q,\pm}(\kappa_n \rho)}{\partial \rho}\right|_{\rho=R}.$$

Dispersion relation can be found from Eq. (9), by solving equation $\det(\hat{D}) = 0$. It is a transcendental equation for $k$ and it has a set of solutions (frequencies $\omega$) for each choice of the index *m*. Taking into account relation for transversal and longitudinal wave numbers (3), we can obtain an implicit relation between frequency $\omega$ and wave number $k$ in the magnetic nanowire. For each value of Bessel function of order *q*, we may obtain set of several frequencies. The frequencies for the modes differencing in *q* may coincide with each other.

### 2.2. Numerical calculations

Numerical calculations were done with the aid of COMSOL Multiphysics software which uses the Finite Element Method (FEM). To find the magnetization dynamics we solved the linearized LLE for Ni nanowire of the parameters stated above (Sec. 2.1). We made the same assumptions concerning the magnetic configuration and considered the same terms in the effective magnetic field as in the analytical model (Sec. 2.1). The Cartesian coordinate system was selected as a default one for numerical calculation performed with the aid of the software we used.

The LLE (1) for the magnetization vector **M** can be linearized in the form of a set of two differential equations for complex amplitudes of dynamical components of magnetization $m_x$ and $m_y$:

$$i\omega m_x(\mathbf{r}) = \gamma\mu_0 \left[ H_0 m_y(\mathbf{r}) - \alpha(\mathbf{r})\left(\nabla_\perp^2 m_y(\mathbf{r}) - k^2 m_y(\mathbf{r})\right) \right.$$
$$\left. + M_0(\mathbf{r})\left(\frac{\partial}{\partial y}\varphi(\mathbf{r})\right) \right], \qquad (11)$$

$$i\omega m_y(\mathbf{r}) = \gamma\mu_0 \left[ -H_0 m_x(\mathbf{r}) + \alpha(\mathbf{r})\left(\nabla_\perp^2 m_x(\mathbf{r}) - k^2 m_x(\mathbf{r})\right) \right.$$
$$\left. - M_0(\mathbf{r})\left(\frac{\partial}{\partial x}\varphi(\mathbf{r})\right) \right], \qquad (12)$$

where $\nabla_\perp^2 = \frac{\partial^2}{\partial x^2} + \frac{\partial^2}{\partial y^2}$. The second term on the right-hand side of Eqs. (11) and (12) have an exchange origin and results directly from the Eq. (2). The exchange constant $A(\mathbf{r})$ and magnetization saturation $M_0(\mathbf{r})$ take the values 0 and $A_{\text{Ni}}$, $M_{0,\text{Ni}}$ outside the nanowire ($\rho > R$) and inside the nanowire ($\rho < R$), respectively.

For considered geometry the static components of the demagnetizing field are equal to zero, nonzero are only the *x*- and *y*-components of the dynamical dipolar field. Using the magnetostatic approximation, the dipolar field can be expressed as a gradient of the scalar magnetostatic potential. With the aid of the Gauss equation, we obtained the following equation which relates magnetization and magnetostatic potential:

$$\nabla^2 \varphi(\mathbf{r}) - k^2 \varphi(\mathbf{r}) - \frac{\partial m_x(\mathbf{r})}{\partial x} - \frac{\partial m_y(\mathbf{r})}{\partial y} = 0. \qquad (13)$$

The Eq. 13 can be used to find dynamic components of demagnetizing field implemented already in Eqs. 11–12, i.e., the last terms on the right-hand side of these equations.

The linearized LLE in the form of the eigenvalue problem (11–12) coupled with Eq. (13) was solved numerically.

## 3. Results

Fig. 1a presents frequency $f = \frac{\omega}{2\pi}$ dependence on a wave number $f(k)$. It was obtained from the semianalytical calculation by the solution of $\det(\hat{D}) = 0$ (see Eqs. 9-10), taking the modes corresponding to the quantum numbers $q = 1$ and $q = 2$ (red solid lines), and from the numerical calculation with the use of FEM (black dots). The numerical calculations confirm the results obtained from the analytical model.

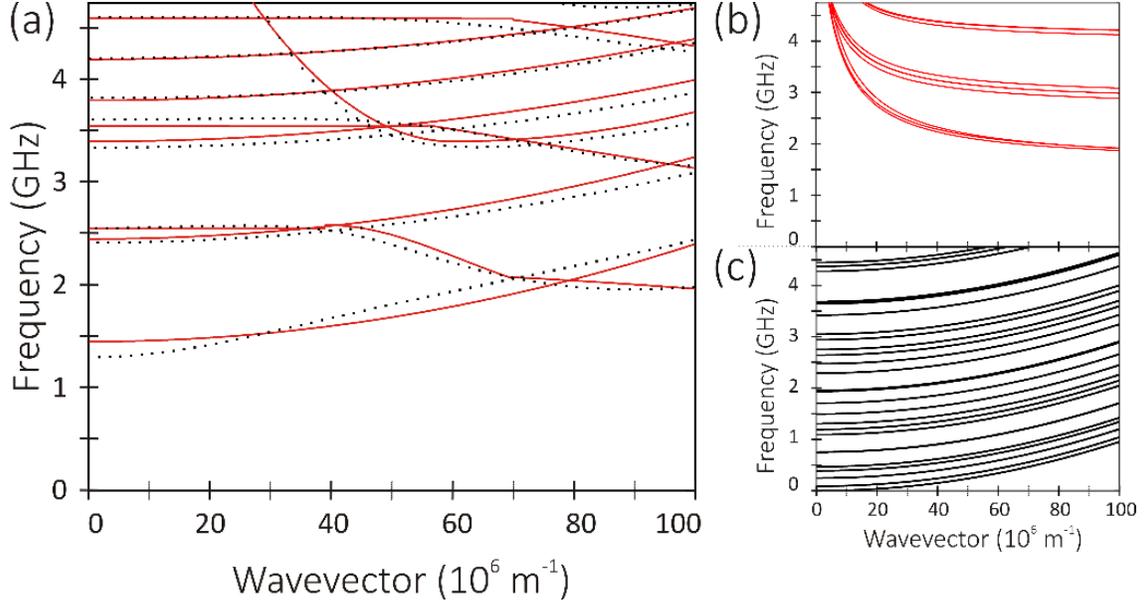

Fig. 1. (a) Dispersion relation of spin waves in the dipolar-exchange regime, propagating along the Ni nanowire obtained by two methods: by numerical calculations (black dots) and by semi-analytical calculations (red lines). (b) Dispersion relation of spin waves with only dipolar interactions included calculated semi-analytically. (c) Dispersion relation of spin waves with only exchange interaction taken into account, calculated using the numerical method. Calculations have been made for Ni nanowire of 60 nm radius and with material parameters: exchange constant $A = 7.46 \times 10^{-13}$ J/m, saturation magnetization: $M_0 = 0.48 \times 10^6$ A/m, the gyromagnetic ratio: $\gamma = 193.6 \frac{\text{GHz}}{\text{T}}$, without any influence of an external magnetic field $H_0 = 0$.

The frequency dependence $f(k)$, shown in Fig. 1a, has a different character for different dispersion branches. Two families of dispersion branches can be distinguished: the first one, for which the frequency decreases with increasing wavenumber which is equivalent to negative group velocity, and the second one, for which this relation is opposite (see e.g. the bands No. 1, 2, 4, 5, counted with increasing frequency at $k = 0$, which belong to the second mentioned family of dispersion branches characterized by positive group velocity). The bands from these two groups crosses and anti-crosses each other. We attribute the negative slope of the first family of the dispersion branches to the impact of dipolar interaction. Due to the magnetostatic shape anisotropy, the static magnetization is oriented along the wire. Therefore the direction of spin wave propagation is here parallel to the static magnetization vector. The same geometrical relation between the direction of wave vector and the direction of static magnetization is observed for so called 'backward volume magnetostatic modes' in magnetic films (named this way because of the negative slope of dispersion branches). For these systems: longitudinally magnetized wire and in-plane magnetized layer, the dynamic magnetization **m** (and its component normal to the surface) precess (oscillate) in-phase for $k = 0$. This dynamic configuration, where the components of **m** normal to the surface oscillate in phase, has large dipolar energy which rises the frequency of the spin wave. The dynamic stray field and magnetostatic potential, produced by the sequence of collaterally arranged dynamic magnetic moments **m**, are enhanced and stretch outside the structure, penetrating deeply the non-magnetic surrounding. For $k \neq 0$ normal component of dynamic magnetization change the phase of oscillation along the wire, which in turn decrees the dipolar energy and reduce the frequency of the spin waves.

The dipolar character of the branches with a negative slope is shown in Fig. 1b. We plotted here the spin wave dispersion for the model of the magnetic nanowire without exchange interaction taken into account. All of the presented dispersion branches have here negative group velocity. Fig. 1c depicts $f(k)$ dependence in the nanowire for only exchange interaction taken into account. The dispersion branches have here always the parabolic shape and are characterized by positive group velocity. Therefore, it nicely confirms, that the dispersion branches of positive (or negative) slope in Fig. 1a correspond to the modes for which the contribution of exchange (or dipolar) energy, related to the spin wave dynamics is dominating. It is also worth to notice that the relative strength of dipolar interactions and exchange interactions changes with the wave number. In Fig. 1a, where both dipolar and exchange interactions were included, all of the dispersion branches become parabolic with positive group velocity for large wavenumbers, i.e., for $k > 1.1 \times 10^6$ m$^{-1}$ (out of scale in Fig. 1a).

In order to investigate quantitatively the character of modes we can calculate from their spatial profiles the contributions of dipolar and exchange energy densities related to these modes:

$$E_{\text{ex}} = \frac{\mu_0 \alpha}{4}\left[\frac{\partial m_x}{\partial x}\frac{\partial m_x^*}{\partial x} + \frac{\partial m_x}{\partial y}\frac{\partial m_x^*}{\partial y} + \frac{\partial m_y}{\partial x}\frac{\partial m_y^*}{\partial x} + \frac{\partial m_y}{\partial y}\frac{\partial m_y^*}{\partial y} + k^2 m_x m_x^* + k^2 m_y m_y^*\right], \tag{14}$$

$$E_{\text{dip}} = \frac{1}{2}\left[Re\left(-\frac{1}{2}\mu_0\left(-\frac{\partial \varphi}{\partial x}m_x^*\right)\right) + Re\left(-\frac{1}{2}\mu_0\left(-\frac{\partial \varphi}{\partial y}m_y^*\right)\right)\right]. \tag{15}$$

These equations will be used in the further part of the paper.

We have already noticed from the Fig. 1a, that the dipolar mode of negative group velocity crosses or anti-crosses with different exchange-dominated modes. Hybridization (anti-crossing) for two modes is possible if their overlap integral:

$$I \sim \int_S (m^p)^* m^l dS \tag{16}$$

is different from zero. The symbols $m^p$ and $m^l$ denote the complex amplitude of the same component of dynamical magnetization for $p^{th}$ and $l^{th}$ mode. The symbol $S$ is the area of the cross-section of the wire. Using expressions (6) for the dynamic magnetization and substituting them into Eq. (16) it is possible to calculate strictly the overlap integral $I$. The hybridization is possible only for the modes form the same set, characterized the same radial wave numbers $\kappa_{n_1} = \kappa_{n_2}$.

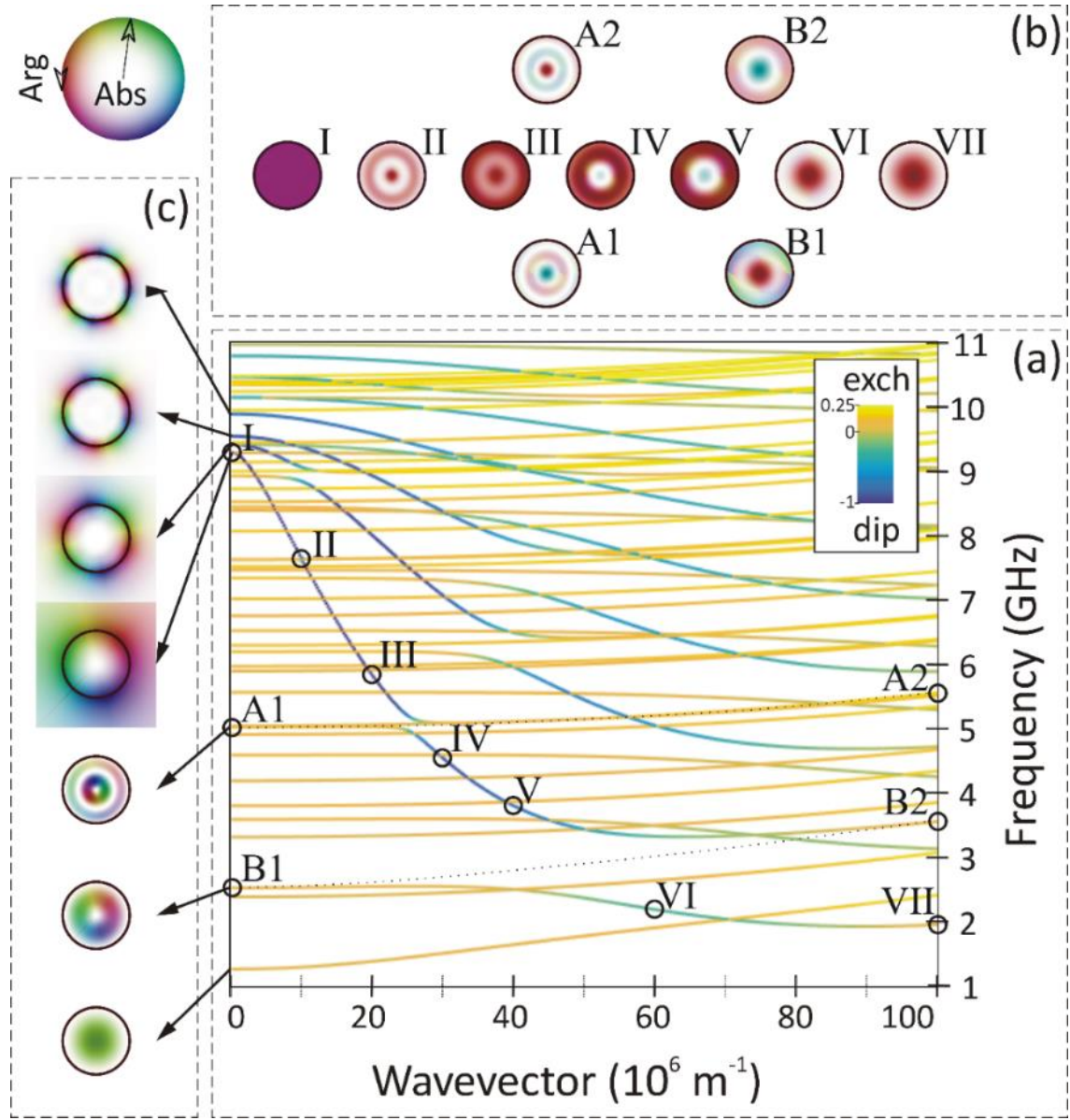

Fig. 2 (a) Dispersion relation of cylindrical Ni nanowire magnetized along its axis in the absence of an external field $H_0 = 0$ calculated using finite element method. (b) Dynamical magnetization profiles for spin wave modes; (c) the profiles of magnetostatic potential, at $k = 0$, for spin wave modes; the inset in the top left corner shows the color scale used to present the amplitude and phase for the profiles of dynamical magnetization and magnetostatic potential. The color of dispersion branches marks the relative contribution of dipolar energy (blue) and exchange energy (yellow), related to the magnetization dynamics. In the set of dispersion branches, for which dipolar interactions are dominating (blue-green lines), we identified at $k = 0$ the fundamental mode I and for $k > 0$ the other mode originating from fundamental mode II - VII. All of them are characterized by an in-phase spin wave precession in the whole cross-section of the wire (see dynamical magnetization profiles I -VII at the top of the figure). The dipolar modes (blue-green lines) anti-cross with the exchange modes (yellow-orange lines) of specific symmetry. We analyzed the anti-crossing of the fundamental mode with two exchange modes: A1-A2 and B1-B2 (see profiles in (b)). The profiles of magnetostatic potential (placed in the column on the left (c)) show the difference in the stray field for dipolar modes and exchange modes.

To discuss in details the features of spin wave dispersion relation in cylindrical wire and to estimate the relative strength of dipolar and exchange interactions, we plotted in Fig. 2 the dependence $f(k)$ in wider frequency range and marked the relative difference of exchange and dipolar energy $(E_{\text{exch}} - E_{\text{dip}})/(E_{\text{exch}} + E_{\text{dip}})$ (calculated according to Eq. 14–15) by coloring the dispersion branches. The colors allow to distinguish the wave dispersion branches for dipolar and exchange dominated modes. All of the negative slope dispersion branches are blue – which means that they are dipolar-dominated, while all of the positive slope dispersion branches are yellowish – which points the exchange-dominated character. The coloring of dispersion branches helps to notice where the hybridization (anti-crossing) of exchange-dominated and dipolar-dominated modes occurs in the dispersion plot.

In experimental studies of spin wave dynamics, the fundamental mode (FM) gives the strongest response for measurements techniques such as ferromagnetic resonance and magneto-optical Kerr effect. To find the fundamental excitation, we have been searching for the mode of homogeneous dynamical magnetization profiles in the wire cross-section and uniform precession along the wire ($k = 0$). The uniform excitation is found at 9.29 GHz, it has 34 ordinal number (counted at $k = 0$), and due to in-phase precession in the whole volume of the nanowire, we will identify this mode as FM. The FM is marked in the dispersion relation presented in Fig. 2a by Roman numeral I. In Fig. 2a and Fig. 2b the evolution of the mode originating from FM (MOFFM) is presented in dependence on the wave number $k > 0$. The dynamical magnetization profiles and the corresponding values of frequency and wave number in dispersion plot are marked by Roman numerals I-VII. The MOFFM is strongly dipolar-dominated. Its dynamical magnetization is in phase in the cross-section of the nanowire, although it changes the amplitude of precession in the radial direction after hybridizations with exchange-dominated modes (see Fig. 1b and Fig. 1c where the color means a change of the phase, while the color intensity means the strength of excitation). We can see that the FM hybridize with other modes at least 3 times, but 2 of them are more visible:

- The MOFFM for $k = 20 \times 10^6$ m$^{-1}$, marked as the Roman numeral III at the dispersion relation Fig. 1a and at Fig. 1b, at which the profile of this mode is presented, hybridizes with exchange-dominated mode marked as A1 (which profile is presented in Fig. 1b); after this hybridization the mode III changes into mode IV – before hybridization the greatest concentration of amplitude could be found at the surface of the nanowire and at its center, while after hybridization (at $k = 35 \times 10^6$ m$^{-1}$) the biggest concentration of spin wave amplitude is observed near to the surface of the nanowire. The mode A1 is evolving into mode A2, after which the phase of precession is flipped.

- The MOFFM for $k = 40 \times 10^6$ m$^{-1}$, marked as the Roman numeral V at the Fig. 1a,b hybridizes with exchange-dominated mode marked as B1; the mode V shows the biggest amplitude at the surface and remains without spin wave excitation at the center of nanowire; the B1 mode has got the largest spin wave amplitude at the center and a smaller one at the boundary of the nanowire, at which it changes its phase; after the hybridization the mode VI emerged as MOFFM and loses the strong excitation at the boundary, but gains the amplitude at the center, while the exchange-dominated mode B2 does not show the change of spin wave phase at the boundary in reference to the mode B1.

After those hybridizations, the MOFFM is still homogeneous in phase, with the strongest excitation placed at the centre of the nanowire, which slowly decreases radially in the vicinity of the surface. We have marked by small black dots the probable shape of the dispersion branches of exchange modes which would be valid in the absence of interactions between them and MOFFMs.

In Fig. 2c we have presented the profiles of magnetostatic potential for spin wave modes at $k = 0$. It allows us to explain why the particular modes anti-cross with MOFFM. The anti-crossing modes need to have similar magnetostatic potential symmetry in the nanowire region. At the bottom, we have presented the magnetostatic potential for the first mode (of the lowest frequency). It is homogeneous in phase and its amplitude (confined in the magnetic material) is concentrated at the center of the nanowire. Then, the magnetostatic potential for the exchange-dominated modes (A1 and B1) of higher frequencies is plotted. These modes hybridize (for larger $k$) with MOFFM modes. For the modes A1 and B1, the magnetostatic potential is also concentrated only in the nanowire region but is less homogeneous than for the lowest mode. Its amplitude changes more in the radial direction (node in the center of the wire –A1 and B1, and additional circular node line between center and surface – B1). The phase of magnetostatic potential for modes A1 and B1 is non-uniform – it changes by $2\pi$ in the azimuthal direction (around the axis of the wire). Finally, we present the profile of magnetostatic potential for FM mode and for the higher dipolar-dominated modes. The magnetostatic potential for those modes is mainly concentrated near to the surface of the nanowire and penetrates the air outside. The magnetostatic potential extends into the air far away from the surface of the nanowire and while circulating around the nanowire it changes its phase by $2\pi$. The A1 and B1 modes are characterized by the same change of the phase around the nanowire (equal $2\pi$). Rest of the exchange-dominated modes, which do not hybridize with MOFFM, have got a different angular distribution of the phase of magnetostatic potential. While the FM has got a change of phase around the nanowire by $2\pi$, the second dipolar-dominated mode is characterized by the change of phase equal $4\pi$, the third mode – $6\pi$, the forth – $8\pi$, etc.

The dipolar-dominated modes are characterized by high group velocity at small wave numbers. It results from the presence of dynamic demagnetizing field (and magnetostatic potential) outside of the wire. This field couples the precessing magnetic moments stronger than the field (and magnetostatic potential) confined mostly in the volume of the wire. The MOFFM is characterized by the largest range of penetration of magnetostatic potential outside of the wire, and this explains why MOFFM has the largest group velocity among the dipolar-dominated modes in this system.

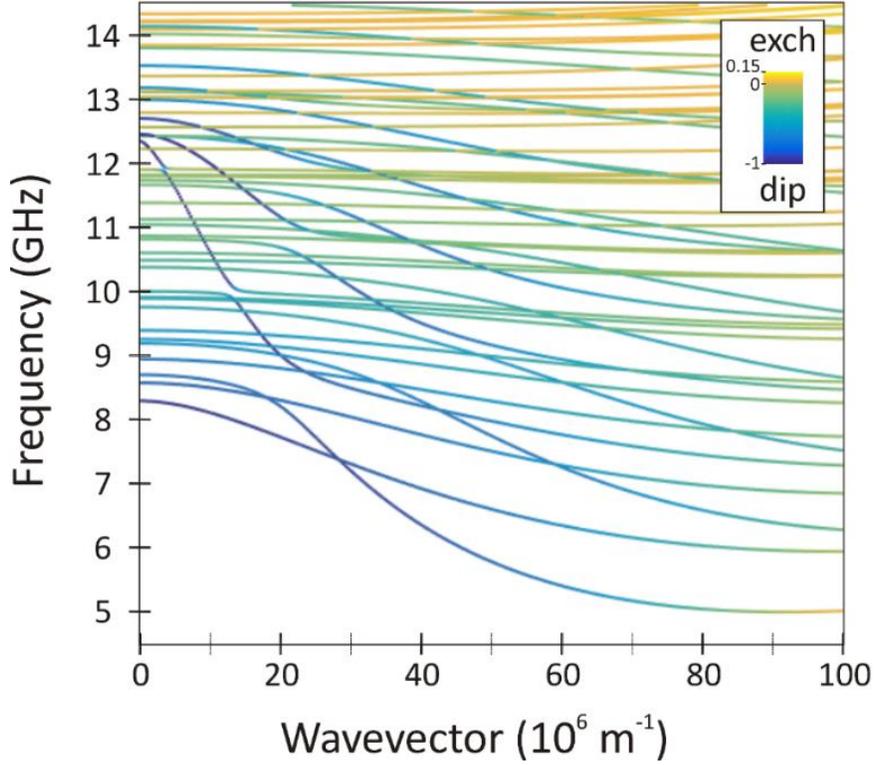

Fig. 3 Dispersion relation of spin waves, in dipolar-exchange regime, propagating along the cylindrical Ni nanowire of 60 nm radius obtained by numerical calculations. The external magnetic field $\mu_0 H_0 = 100$ mT was applied in the direction of nanowire axis. The color of dispersion branches marks the relative contribution of dipolar energy (blue) and exchange energy (yellow) related to magnetization dynamics.

Let us now consider the same system but under the influence of the external magnetic field applied along the axis of the nanowire. We have made calculations for this system using FEM and plotted the dispersion relation (Fig. 3) with branches coloured according to the contribution of exchange (yellow) and dipolar (blue) term to the spin wave energy [calculations are based on Eqs. (14-15)]. We can see that after application of the external magnetic field all dispersion branches at low frequencies and long wavelengths became dipolar-dominated, which is manifested in their negative slope and in the change of the colour toward the blue. Also, the density of modes below the FM increased with respect to the dispersion relation with a zero magnetic field, here there are 24 modes in the range 8.3 – 12.4 GHz (Fig. 3), at $H_0 = 0$ (Fig. 2) there was 34 modes in 1.3 – 9.3 GHz range.

To explain the change in the density of spin wave modes after application of the external field $H_0$, we turned our attention to the dependences of the eigenfrequencies at $k = 0$ on the field $H_0$, which can be different for dipolar and exchange dominated modes.

Indeed, such various dependences were already discussed in Ref. 53. The FM satisfies the Kittel's resonance condition for the frequency of uniform precession in the whole volume of an axially magnetized infinite cylinder:

$$f = \frac{\gamma\mu_0}{2\pi}(H_0 + M_0/2), \qquad (17)$$

while the lowest frequency mode supposed to exhibits the substantially different dependence:

$$f = \frac{\gamma\mu_0}{2\pi}\sqrt{H_0(H_0 + M_0)}. \qquad (18)$$

The modes of the lowest frequency (below the frequency of FM) and small $k$ show the precession with a significantly reduced radial component of the dynamic magnetization (see Figs. 4 and 5). We can notice, that these modes (called further circumferential modes (CMs) [53]) show some similarities to the FM in a magnetic film, where a strongly elliptical precession also is observed. In the nanowire both the tangential and normal components of the dynamic magnetization precess in phase for the lowest CM (see the first column of the Figs. 4 and 5), which also resembles spin wave dynamics of the FM in a planar geometry. Based on this argument, we can justify the use of the same formula for the description of the dependence $f(H_0)$ for FM in the film and the CM of the lowest frequency in the nanowire investigated in our study (see the first column of the Figs. 4 and 5).

The FM (the fourth column in Figs. 4 and 5) and the other higher modes (the fifth column in Figs. 4 and 5), generating the strong dynamical dipolar field outside the nanowire, are dipolar-dominated. The CMs, on the other hand, minimize the dynamical stray field outside the magnetic structure. It is achieved by the reduction of the normal component of the dynamical magnetization field which is equivalent to the formation of the 'flux closure' configuration of the dynamical magnetization. This intuitively explains the exchange-dominated character of CM.

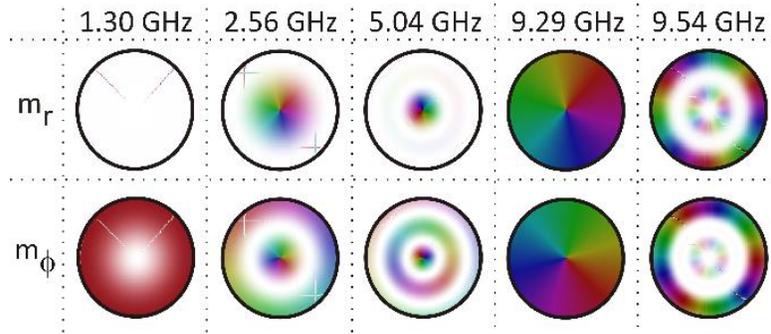

Fig. 4 Radial and azimuthal components of dynamical magnetization: $m_r$ and $m_\varphi$, for spin wave modes (marked in the Fig. 2 by empty circles) in the absence of external magnetic field $H_0 = 0$, at $k = 0$. The phase is presented as a specific color, while the amplitude is shown as the intensity of the color.

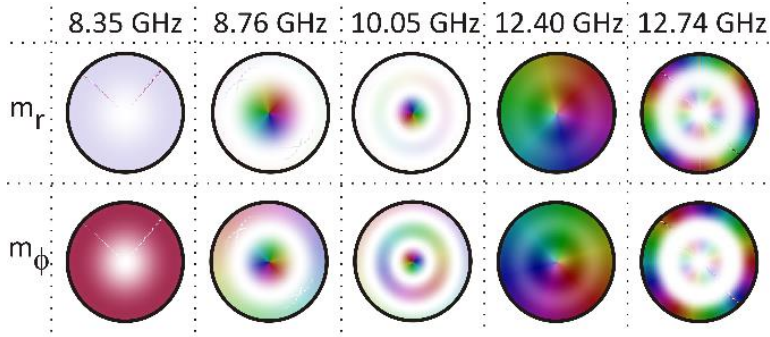

Fig. 5 Radial and azimuthal components of dynamical magnetization: $m_r$ and $m_\varphi$, for spin wave modes (marked in the Fig. 3 by empty circles) in the presence of external magnetic field $\mu_0 H_0 = 100$ mT, at $k = 0$. The phase is presented as a specific color, while the amplitude is shown as the intensity of the color.

Reassuming, the dependences $f(H_0)$ for the dipolar-dominated modes (FM and higher frequency modes in Fig. 2a) and for exchange-dominated modes (yellow lines in Fig. 2a) shall be different as described by Eqs. (17) and (18), respectively. The eigenfrequency of FM increases linearly with the field, shifted up by frequency independent term $\frac{\gamma\mu_0}{2\pi}M_0/2$ in Eq. (17), whereas the frequency of CM modes shall rise faster than linearly with the field. The significant group of CMs modes have initially for $H_0 = 0$ lower frequencies than FM, but with an increase of the field, the exchange-dominated CMs will shift up more in frequency scale than FM (and the other dipolar-dominated modes) and finally can go over FM at the sufficiently high magnetic field.

To check if those predictions match with our computations, we compared the frequencies of FM and CM obtained from the formulas (17) and (18) with the corresponding numerical values readout from the dispersion at $k = 0$ shown in Fig. 2 and Fig. 3, obtained from Eqs. (11-13). In the absence of the external field $H_0$, the frequency of FM takes partially the same value: 9.29 GHz both for the formula (17) and for the numerical calculations. For the field $\mu_0 H_0 = 100$ mT, we obtained the values: 12.38 GHz (from Eq. 17) and 12.40 GHz (from Fig. 3), which means that FM is pushed up in frequency for about 4 GHz after application of the field. According to Eq. (18) the frequency of CM goes to zero in the limit $H_0 \to 0$ but our numerical value (Fig. 2) is nonzero: 1.30 GHz. This upward shift of the frequency for the numerical solution of CM can be attributed to the effect of confinement – the analytical formula (18) does not take into account finite radius of the wire. The application of the external magnetic field $\mu_0 H_0 = 100$ mT shifts the frequency of the CM about 7 GHz up, which is significantly larger than for FM. The frequency for CM reaches here the value: 8.17 GHz (according to Eq. 18) and 8.35 GHz (in Fig. 3). The good match of the simple model predictions with the numerical results justify the hypothesis that the frequencies of dipolar-dominated modes increases faster than the frequencies of the exchange-dominated modes after application of the external magnetic field along the wire.

To clarify the enhancement of the dipolar character of the low-frequency modes, with the increase of the magnetic field, we will discuss the changes in their spatial profiles shown in Fig. 4 and 5. The CM – the mode of the lowest frequency, found at $k = 0$ in Figs. 2 and 3 is exchange-dominated. This mode has a dominant amplitude in the azimuthal direction, its radial dynamical magnetization in the cross-section of the wire is very weak (see the first column in Figs. 4 and 5). Due to that the CM mode practically do not show dynamic magnetostatic potential outside of the nanowire (see Fig. 2c). Moreover, the profile of CM mode indicates that the dynamical dipolar field has flux closure structure and as a result weakly couples the precessing magnetic moments. The increase of applied magnetic field reduces slightly the difference between the azimuthal and radial component of dynamical magnetization. It means that CM gains, to a small extent, dipolar character.

The next two selected exchange-dominated modes are presented in the 2$^{nd}$ and 3$^{rd}$ columns in Figs. 4 and 5. These are the modes that hybridize with the FM mode for the larger $k$ wavenumbers (see Fig. 2 and Fig. 3). They are marked as empty circles at $k = 0$ on the frequency axis in Figs. 2 and 3. The magnetostatic potential for these modes is also concentrated mostly inside the nanowire, which means there is practically no dynamic dipolar field outside of the nanowire (see Fig. 2c). For these modes, the radial component of dynamical magnetization appears in the center of the nanowire, although still, it is not present at the edge of the nanowire. The azimuthal component of dynamical magnetization is concentrated both next to the edge of the nanowire and at the center of the nanowire, with one node at the radial direction for the case of the mode B1 and with two nodes in the case of the mode A1. The profile of dynamical magnetization close to the surface (small $m_r$ and nonzero $m_\theta$) ensures the reduction of dynamical demagnetizing field outside of the nanowire and is related to the exchange character of this mode. The increase of applied magnetic field makes the difference between $m_r$ and $m_\theta$ slightly less noticeable which is manifested in the increase of dipolar character of these modes (compare Figs. 2 and 3).

The FM (the fourth column in Figs. 4 and 5 and mode marked in Figs. 2 and 3 by roman I) is the mode precessing homogeneously in the entire cross-section of the nanowire. It can be seen in Fig. 2 where the amplitudes and phases of $m_x$ and $m_y$ component of dynamical magnetization are constant (see the profile No. I). The $m_r$ and $m_\theta$ are the components in the polar coordinate system and they have equally large amplitude through the whole cross-section of the nanowire, but they are shifted in the phase of $\pi/2$. The phase of m$_r$ and m$_\theta$ is changing by $2\pi$ on the whole perimeter of the nanowire which is also the consequence of homogeneity of FM observed in this curvilinear system - as the phase changes by $2\pi$, the dynamic magnetization $\boldsymbol{m} = \hat{r}m_r + \hat{\theta}m_\theta$ rotates homogeneousely in the whole cross-section of the wire by the same angle: $2\pi$ (like single magnetic moment). This produces the strong dynamical stray field outside of the wire.

In the last column of Figs, 4 and 5 the further dipolar mode is presented (i.e. the second dipolar mode above the FM mode). It has got quite large magnetostatic potential (dipolar magnetic field) outside of the nanowire region (see the Fig. 2c, the highest empty circle mode). For this mode the phase of the magnetostatic potential changing of 6π around the perimeter of the nanowire. It has got nodes at the centre of the nanowire and at the middle between the edge and the centre. Both components $m_r$ and $m_\varphi$ have got the same large amplitude in the middle of the nanowire (surrounding the centre) and at the edge of the nanowire. The phase is changing at the circumference of the nanowire of 6π, which is the same amount as for the magnetostatic potential. This mode of dynamical magnetization gives the same dipolar field as the one coming from the magnetic moments rotated by 2/3π in respect to each other.

The application of external field does not affect significantly dipolar modes. It is manifested by almost unaltered ratios between the amplitudes of $m_r$ and $m_\varphi$ components on the surface of the nanowire.

## 4. Conclusions

We calculated the spin wave dispersion relation in cylindrical nanowire magnetized along its axis, taking into account both dipolar and exchange interactions. The dipolar or exchange character of particular mode was identified by the calculation of exchange and dipolar energy of the corresponding spin wave profile. In the rich spectrum of eigenmodes, we were able to find the dipolar-dominated modes (characterized by the large negative values of group velocity at small wave numbers) and the exchange modes (of positive group velocity).

The dispersion branches of the modes which have positive and negative slopes (different signs of group velocities) cross or anti-cross with each other. The anticrossing occurs when the modes have the same symmetry of magnetostatic potential (i.e. with the same number of nodes for the phase of the magnetostatic potential in one circulation around the axis of the wire). The detailed studies of anticrossing were done for the modes of the largest contribution of dipolar energy, originating from the fundamental mode.

We present also the study of the influence of external magnetic field on spin wave dispersion relation in the magnetic nanowire. After application of the magnetic field along the wire, the dispersion relation changes noticeably – most of the modes of small wavenumbers are then characterized by negative group velocity and exhibit stronger dipolar character. The frequencies of the modes are shifted upwards but this shift is more significant for the exchange-dominated modes. This mechanism explains why more dipolar-dominated modes occupy the bottom of the spin wave spectrum after application of magnetic field.

## 5. Acknowledgements

This project has received funding from the European Union's Horizon 2020 research and innovation programme under the Marie Skłodowska-Curie grant agreement No 644348 (MagIC) and from the financial resources of MNiSW for science in the years 2017-2019 granted for the implementation of an international co-financed project. J.W.K. would like to acknowledge the support of the Foundation of Alfried Krupp Kolleg Greifswald. J.R. would like to acknowledge the financial support from the Adam Mickiewicz University Foundation and from the National Science Centre of Poland under ETIUDA grant number UMO-2017/24/T/ST3/00173, work at Argonne was supported by U.S. Department of Energy, Office of Science, Office of Basic Energy Sciences, under contract no. DE-AC02-06CH11357.

The authors also gratefully thank professor R.I. Peleschak for the usefull ideas.